\documentclass[utf8]{FrontiersinHarvard} 

\usepackage{url,hyperref,lineno,microtype,subcaption,nicefrac}
\usepackage[onehalfspacing]{setspace}

\newcommand{\Ni}{\ensuremath{^{56}\mathrm{Ni}}\xspace}

\newcommand{\Msun}{\,\ensuremath{\mathrm{M}_\odot}\xspace}

\newcommand{\Zsun}{\,\ensuremath{\mathrm{Z}_\odot}\xspace}

\newcommand{\Msunpyr}{\,\ensuremath{\Msun~\mathrm{yr^{-1}}}\xspace}
\newcommand{\kmps}{\,\ensuremath{\mathrm{km~s^{-1}}}\xspace}

\usepackage{xspace}

\def\keyFont{\fontsize{8}{11}\helveticabold }
\def\firstAuthorLast{Moriya {et~al.}} 
\def\Authors{Takashi J. Moriya\,$^{1,2,3,*}$, Steve Schulze\,$^{4}$, Alexander Heger\,$^{3}$, Sergei I. Blinnikov\,$^{5,7}$ and Marat Sh. Potashov\,$^{6,7}$}
\def\Address{$^{1}$National Astronomical Observatory of Japan, National Institutes of Natural Sciences, Tokyo, Japan \\
$^{2}$Graduate Institute for Advanced Studies, SOKENDAI, Tokyo, Japan \\
$^{3}$School of Physics and Astronomy, Monash University, Clayton, VIC, Australia \\
$^{4}$Department of Particle Physics and Astrophysics, Weizmann Institute of Science, Rehovot, Israel\\
$^{5}$NRC Kurchatov Institute, Moscow, Russia\\
$^{6}$Keldysh Institute of Applied Mathematics, Russian Academy of Sciences, Moscow, Russia\\
$^{7}$Sternberg Astronomical Institute, Moscow State University, Moscow, Russia
}
\def\corrAuthor{Takashi J. Moriya}

\def\corrEmail{takashi.moriya@nao.ac.jp}

\begin{document}
\onecolumn
\firstpage{1}

\title[CSM interaction in SN~2018ibb]{Properties of the circumstellar matter around the pair-instability supernova candidate SN~2018ibb revealed by its light curve} 

\author[\firstAuthorLast ]{\Authors} 
\address{} 
\correspondance{} 

\extraAuth{}

\maketitle

\begin{abstract}
SN~2018ibb is one of the best pair-instability supernova (PISN) candidates identified to date. It, however, showed unexpected blue flux excess in late-phase spectra, likely originating from the interaction between supernova (SN) ejecta and dense circumstellar matter (CSM). We develop synthetic light-curve models of PISNe interacting with dense CSM and estimate the CSM properties of SN~2018ibb by comparing the synthetic and observed light curves. We found that the bolometric luminosity evolution of SN~2018ibb from $150\,\mathrm{d}$ after the peak can be well reproduced by the interaction with the CSM formed by 
a mass-loss rate of 
$0.01\,(v_\mathrm{CSM}/1{,}000\,\kmps)\,\Msunpyr$, where $v_\mathrm{CSM}$ is the CSM velocity.
The observed luminosity break at around $300\,\mathrm{d}$ from the peak indicates that the CSM interaction ended at this time and the dense CSM radius was $8.5\times 10^{16}\,\mathrm{cm}$. The CSM radius then implies that the mass loss of the progenitor was enhanced for $28\,(v_\mathrm{CSM}/1{,}000\,\kmps)^{-1}\,\mathrm{yr}$ before explosion. We conclude that PISN progenitors may experience short-term mass-loss enhancement within decades before explosion, similar to what is often observed in progenitors of core-collapse SNe.

\section{}


\tiny
 \keyFont{ \section{Keywords:} massive stars, supernovae, SN 2018ibb, pair-instability supernovae, mass loss, circumstellar matter} 
\end{abstract}

\section{Introduction}
Pair-instability supernovae (PISNe) are predicted thermonuclear explosions of very massive stars (\citealt{Barkat1967,Rakavy1967}, see \citealt{Renzo2026} for a recent review). When the pair creation in the core of massive stars becomes efficient enough, they become dynamically unstable and start to collapse. The collapse leads to the runaway O burning in the core, resulting in their explosion. The required core mass to explode as PISNe is estimated to be between $\simeq 65~\Msun$ and $\simeq 135~\Msun$ \citep[e.g.,][]{Heger2002}, although the exact mass range depends on the C fraction in the core \citep[e.g.,][]{Takahashi2018,Farmer2019,Kawashimo2024}. The corresponding zero-age main sequence mass is between $\simeq 150~\Msun$ and $\simeq 300~\Msun$ if mass loss and rotation are insignificant (e.g., \citealt{Hirschi2025}, see also \citealt{Chatzopoulos2012,Marchant2020} for the effects of rotation). Mass loss, however, is large in massive stars \citep[e.g.,][]{Vink2022}, preventing them from keeping the massive core required to explode as PISNe. Whereas the mass loss from massive stars is expected to be small in metal-poor environment, it becomes significant as metallicity becomes higher.  \citet{Heger2003} predicted that PISNe can only exist at low metallicity and \citet{Langer2007} estimated a threshold of around $\Zsun/3$, but this metallicity threshold is uncertain because of the uncertain metallicity dependence of wind mass loss \citep{Sabhahit2023}. Unknown mass loss such as eruptive mass loss can also affect the threshold.

The event rate of PISNe in the local ($z\lesssim 0.3$) Universe is expected to be low because the present-day average metallicity is high \citep{Briel2022,Briel2024,Tanikawa2023,Gabrielli2024,Simonato2025}. Yet, owing to intensive modern transient surveys, a couple of nearby PISN candidates have been discovered so far \citep[e.g.,][]{GalYam2009,Terreran2017,Hiramatsu2026}. Among them, SN~2018ibb is one of the best PISN candidates identified so far \citep{Schulze2024}. SN~2018ibb is a Type~Ic SN with the rise time of more than $93\,\mathrm{d}$ and peak luminosity of around $2\times 10^{44}~\mathrm{erg~s^{-1}}$, consistent with the predicted PISN light curves \citep{Kasen2011,Dessart2013,Whalen2014,Kozyreva2014,Kozyreva2017,Gilmer2017}. Its overall spectroscopic properties also matched the PISN prediction \citep{Jerkstrand2016} including strong [O$\,$\textsc{iii}] \citep{Chugai2024}. In the late-phase spectra of SN~2018ibb, however, flux excess below around $5\mathord,500\,\mathrm{\AA}$, which is not predicted to exist in PISNe, was identified. The blue flux excess was suggested to originate from the interaction between PISN ejecta and circumstellar matter (CSM) surrounding the progenitor \citep{Schulze2024}. The necessity of an extra luminosity source in addition to the radioactive decay of \Ni was also suggested by modeling the light curve \citep{Kozyreva2024,Chugai2024lc,Nagele2024}.

In this paper, we develop the PISN light-curve models with CSM and compare them with that of SN~2018ibb. By identifying the effects of the CSM in the light curve of SN~2018ibb, we estimate the CSM properties surrounding the progenitor of SN~2018ibb and discuss their implications in the PISN progenitor evolution. In Section~\ref{sec:method}, we introduce our methods of PISN light-curve modeling with CSM interaction. We present our results in Section~\ref{sec:results} and discuss them in Section~\ref{sec:discussion}.

\section{Methods}\label{sec:method}
We adopt the PISN model from Moriya et al. (in preparation). They presented a grid of PISN models of bare He stars with various metallicities covering from primordial to super-solar metallicities. The PISN progenitor masses range from $\simeq 65\,\Msun$ to $\simeq 135\,\Msun$ depending on metallicities. The H layer may have been lost due to stellar winds \citep{Langer2007}, binary interaction (envelope stripping), or not be present due to chemical homogeneous evolution induced by rapid rotation \citep[e.g.,][]{HL00}. Although we adopt He star PISN models, SN~2018ibb did not show any signatures of He. Overall \Ni-powered light-curve properties are, however, likely similar to He star and CO star progenitors as long as the electron-scattering is the dominant opacity source in the ejecta \citep[e.g.,][]{Moriya2020}. The interaction-powered light curves are not affected by the composition, either. Thus, adopting He star PISN progenitor models does not significantly affect our conclusions.

\begin{figure}
 \begin{center}
  \includegraphics[width=10cm]{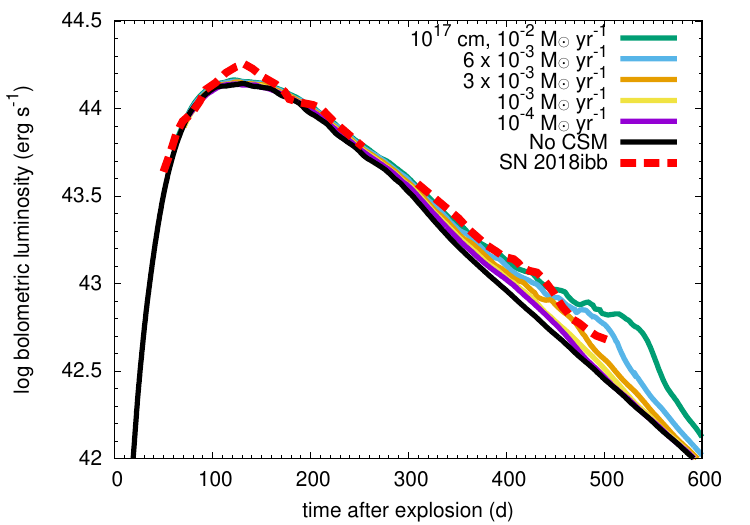} 
 \end{center}
\caption{
PISN light curves with and without CSM. The PISN light curves with CSM in this figure has the fixed radius of $10^{17}\,\mathrm{cm}$. The mass-loss rates range from $10^{-4}\,\Msunpyr$ to $10^{-2}\,\Msunpyr$ ($v_\mathrm{CSM}=1000\,\kmps$). The bolometric light curve of SN~2018ibb is shown for comparison.
}
\label{fig:bolometric_mdot}
\end{figure}

The PISN progenitor evolution and explosion were computed by the \textsc{Kepler} hydrodynamical stellar evolution code \citep{Heger2000,Woosley2002,Heger2005,Woosley2007}. The subsequent PISN light curves are computed by the \texttt{STELLA} multi-frequency radiation hydrodynamics code \citep{Blinnikov1999,Blinnikov2000,Blinnikov2006}, which is briefly explained below.  Based on an estimated metallicity of the host galaxy of $\sim0.25\,\Zsun$, we searched for a PISN model that matches the bolometric light-curve evolution around the peak luminosity of SN~2018ibb among the PISN models with $0.25\,\Zsun$. We found that the $121\,\Msun$ PISN model, which has an explosion energy of $E_\mathrm{ej}=7.5\times 10^{52}\,\mathrm{erg}$ and a \Ni mass of $27.4\,\Msun$, matches the overall bolometric light-curve behavior of SN~2018ibb around the peak luminosity (Figure~\ref{fig:bolometric_mdot}). The explosion energy and the \Ni mass were obtained by the \textsc{Kepler} computation, and they are not free parameters in the PISN explosion. These properties are consistent with the previous estimates by \citet{Schulze2024,Kozyreva2024,Nagele2024}.  The ejecta mass of the model is $M_\mathrm{ej}=121\,\Msun$ because the entire progenitor explodes as a PISN without leaving any compact remnant.  The rise time and peak luminosity of the bolometric light curve of the $121\,\Msun$ model are $128\,\mathrm{d}$ and $1.3\times 10^{44}\,\mathrm{erg~s^{-1}}$, respectively. 

Moriya et al. (in preparation) computed synthetic light curves from $100\,\mathrm{s}$ after collapse. In this work, however, we started light-curve computation from $10\mathord,000\,\mathrm{s}$ after collapse to ease numerical light-curve computations with CSM.  The ejecta homologously expanded at $100\,\mathrm{s}$, and the ejecta continue to expand homologously until $10\mathord,000\,\mathrm{s}$. The expansion until $10\mathord,000\,\mathrm{s}$ was numerically computed by \texttt{STELLA}. The outermost layer of the ejecta was located at $2.8\times10^{14}\,\mathrm{cm}$ at $10\mathord,000\,\mathrm{s}$.  The CSM is attached above this radius at this time. The rise time of the light curve of the $121\,\Msun$ model is $128\,\mathrm{d}$.  Thus, starting numerical computations at $10\mathord,000\,\mathrm{s} \simeq 0.1\,\mathrm{d}$ does not affect the light-curve properties. 

The CSM was attached on top of the ejecta at $10\mathord,000\,\mathrm{s}$ above $2.8\times 10^{14}\,\mathrm{cm}$.  The CSM density structure $\rho_\mathrm{CSM}$ was assumed to be
\begin{equation}
    \rho_\mathrm{CSM} = \frac{\dot{M}}{4\pi v_\mathrm{CSM}}r^{-2}, \label{eq:rhocsm}
\end{equation}
where $\dot{M}$ is the mass-loss rate, $v_\mathrm{CSM}$ is the CSM velocity, and $r$ is the radius. In this work, we assumed $v_\mathrm{CSM}=1\mathord,000\,\mathrm{km~s^{-1}}$, which corresponds to typical terminal wind velocities of Wolf-Rayet stars \citep[e.g.,][]{Hamann2006}, and altered $\dot{M}$ to change the CSM density. 
The CSM velocity may vary on the order of $10-1{,}000\,\kmps$, depending on the mass-loss mechanisms such as binary stripping \citep[e.g.,][]{MacLeod2020,scherbak2025,Ercolino2025,Mandal2026,Wei2026}.
The mass-loss rates in this paper correspond to the case of $v_\mathrm{CSM}=1{,}000\,\kmps$ unless notified, but they depend on the assumed $v_\mathrm{CSM}$, as in Equation~(\ref{eq:rhocsm}).

The synthetic PISN light curves with CSM were computed by using \texttt{STELLA} \citep{Blinnikov1999,Blinnikov2000,Blinnikov2006}. \texttt{STELLA} numerically computes the hydrodynamic equations coupled with the radiative transfer equations by using the variable Eddington factor method. The multi-wavelength radiation transfer was solved by adopting $100$ wavelength bins from $1\,\mathrm{\AA}$ to $5\times 10^4$\,\AA\ uniformly on a log scale. \texttt{STELLA} assumes spherical symmetry, which is likely a reasonable assumption for PISNe \citep{Joggerst2011,chen2014}. \texttt{STELLA} has been adopted to compute synthetic light curves of SNe interacting with CSM \citep[e.g.,][]{Moriya2013a,Moriya2025}. 

\begin{figure}
 \begin{center}
  \includegraphics[width=10cm]{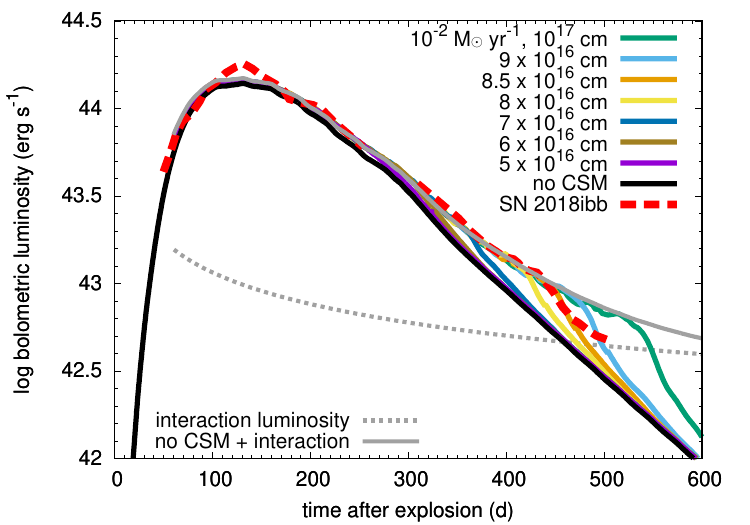} 
 \end{center}
\caption{
PISN light curves with different CSM radii. The mass-loss rate $\dot{M}$ is fixed to $10^{-2}~\Msunpyr$ ($v_\mathrm{CSM}=1000\,\kmps$) in the models with CSM in this figure. The CSM radii range from $5\times 10^{16}~\mathrm{cm}$ to $10^{17}~\mathrm{cm}$. The gray dashed line is an analytic estimate for the luminosity contribution from the CSM interaction. By adding the analytic estimate and the PISN model without CSM, the gray luminosity evolution (the solid line) is obtained.
}
\label{fig:bolometric_radius}
\end{figure}

\section{Results}\label{sec:results}
We first present the results of the PISN light-curve computations with CSM obtained by assuming the CSM radius of $10^{17}\,\mathrm{cm}$ (Figure~\ref{fig:bolometric_mdot}). Within the mass-loss rate range we investigated in this paper ($10^{-4}-10^{-2}\,\Msunpyr$), the CSM interaction has little effect on the light curve near the peak luminosity. After around $300\,\mathrm{d}$ since explosion, however, the synthetic light curves with CSM start to deviate from that without CSM. As the $\dot{M}$ (i.e., CSM density) increases, the luminosity after around $300\,\mathrm{d}$ increases. We found that the luminosity of the model with $\dot{M}=10^{-2}~\Msunpyr$ matches well to that of SN~2018ibb in $300$--$430\,\mathrm{d}$. At around $430\,\mathrm{d}$, SN~2018ibb showed the luminosity break.

Figure~\ref{fig:bolometric_radius} shows the synthetic light-curve models with the same $\dot{M}=10^{-2}\,\Msunpyr$ but with different CSM radii. We find that the luminosity break is determined by the CSM radius, and the luminosity break observed in SN~2018ibb can be reproduced by setting the CSM radius at $8.5\times 10^{16}~\mathrm{cm}$. Assuming the wind velocity of $1\mathord,000\,\kmps$, the mass-loss rate of the progenitor of SN~2018ibb should have increased to $10^{-2}\,\Msunpyr$ from $28\,\mathrm{yr}$ before explosion to set the CSM radius at $8.5\times 10^{16}\,\mathrm{cm}$.
The period of the mass-loss enhancement depends on the assumed $v_\mathrm{CSM}$ as $28\,(v_\mathrm{CSM}/1,000\,\kmps)^{-1}\,\mathrm{yr}$.
After the luminosity break, the synthetic light curves turn back to the synthetic light curve without CSM.  The bolometric light curve of SN~2018ibb, however, showed a flattening after the break.  This flattening may indicate that SN~2018ibb started to interact with less dense CSM after the break.

The CSM luminosity contribution to the bolometric light curves can be simply understood by using an analytic model of the bolometric luminosity evolution of SNe interacting with CSM \citep[e.g.,][]{Moriya2013}. When $\rho_\mathrm{CSM}\propto r^{-2}$, the bolometric luminosity of interacting SNe ($L_\mathrm{int}$) can be expressed as
\begin{equation}
    L_\mathrm{int} = L_1 t^{-\frac{3}{n-2}},
\end{equation}
where\begin{equation}
    L_1 = \frac{\varepsilon}{2}\left(\frac{\dot{M}}{v_\mathrm{CSM}}\right)^{\frac{n-5}{n-2}}\left(\frac{n-3}{n-2}\right)^{3}\left[\frac{2}{(n-4)(n-3)(n-\delta)}\frac{\left[2(5-\delta)(n-5)E_\mathrm{ej}\right]^{(n-3)/2}}{\left[(3-\delta)(n-3)M_\mathrm{ej}\right]^{(n-5)/2}}\right]^{\frac{3}{n-2}},
\end{equation}
$\varepsilon$ is the conversion efficiency from kinetic energy to radiation. In the analytic formula, we assume that the density structure $\rho_\mathrm{ej}$ of the SN ejecta can be approximately by two power-law components, i.e., $\rho_\mathrm{ej}\propto r^{-n}$ outside and $\rho_\mathrm{ej}\propto r^{-\delta}$ inside. We find that our PISN ejecta density structure can be approximated by $n=7$ and $\delta=1$. In Figure~\ref{fig:bolometric_radius}, we present the expected luminosity contribution from the CSM interaction based on the analytic formula (the gray dashed line) as well as the combined luminosity of the PISN luminosity without CSM and the CSM interaction luminosity (the gray solid line). We found that the combined luminosity input reproduces the the overall luminosity evolution of the PISN model with CSM until the luminosity break. Here, we assumed $\varepsilon = 0.2$ which reproduced the luminosity evolution well. We find that the CSM interaction luminosity does not contribute to the luminosity around the peak. Thus, the PISN properties and CSM properties can be independently constrained by using the different phases in the light curve.

\begin{figure}[t]
  \centering
  \includegraphics[width=\linewidth]{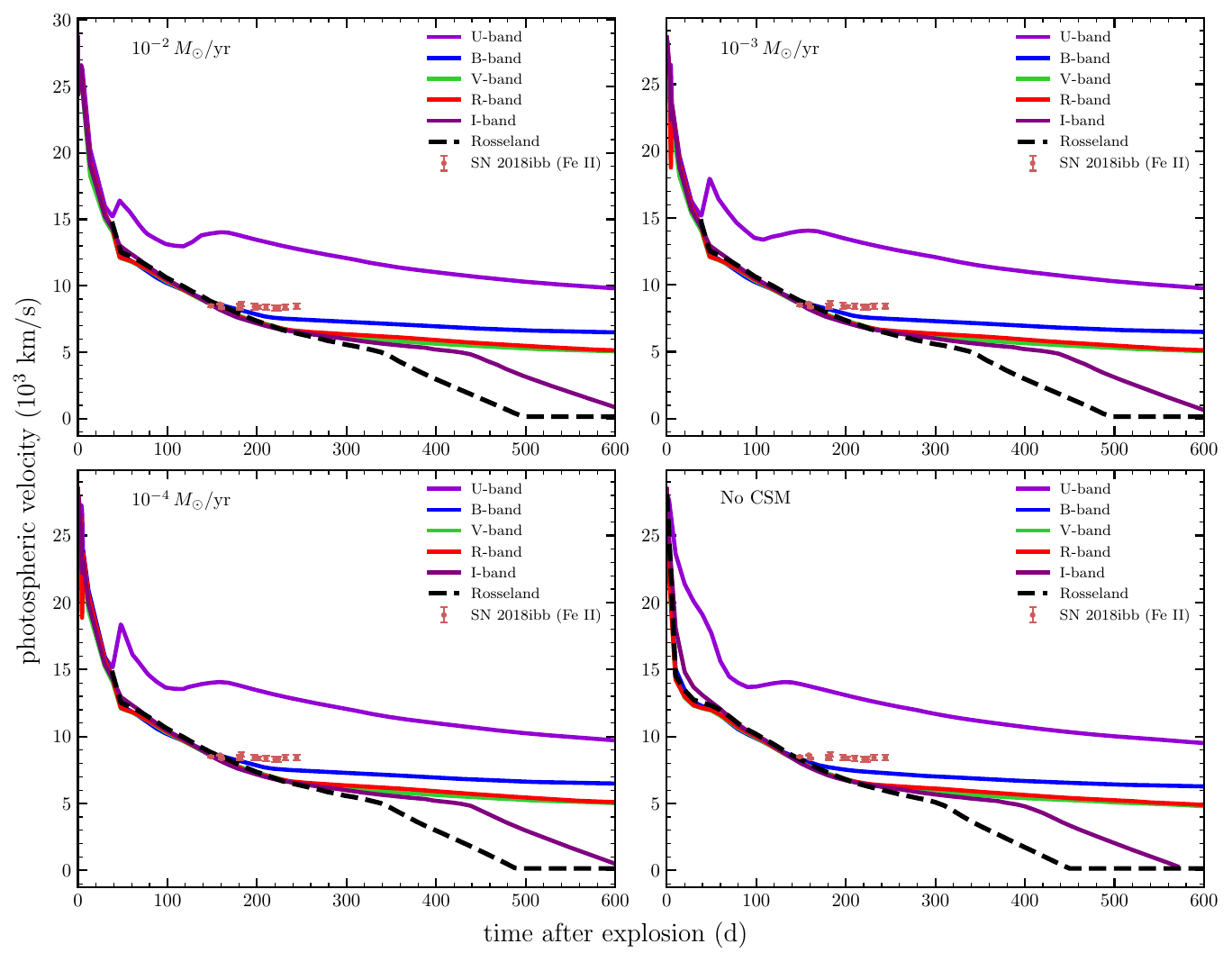} 
  \caption{
  Photospheric velocities of the synthetic models defined
  as the radius where the optical depth in the $U\!BV\!RI$ bands and the Rosseland mean becomes $\nicefrac23$. 
  The different panels correspond to different CSM density. The CSM radius is fixed to $10^{17}~\mathrm{cm}$ in the models in this figure.
  For comparison, the Fe$\,$\textsc{ii} velocity measurements of SN~2018ibb from \citet{Schulze2024} are overplotted.
  The Rosseland photospheric velocity declines with time more steeply than the velocities derived from the $B$ band,
  where the Fe$\,$\textsc{ii} absorption lines used for the velocity
  measurements are located.
}
\label{fig:phv}
\end{figure}

The photospheric velocity evolution of some representative models is presented in Figure~\ref{fig:phv}. We found that the photospheric velocity evolution is not strongly affected by the CSM interaction.
The photospheric velocity inferred from the $B$ band in the our PISN model
is roughly consistent with the Fe$\,$\textsc{ii} absorption-line velocities measured by \citet{Schulze2024}.
The $B$ band is a good approximation because the Fe$\,$\textsc{ii} $\lambda\lambda$4924, 5018, 5169 lines
usually trace the photospheric velocity in the optical,
i.e., the velocity of the layer where the continuum optical depth in the blue/visual range is around unity.

The photosphere lies at different geometric depths at different frequencies.
Caution should be exercised when using the photospheric velocity defined by the Rosseland-mean opacity at $\tau=\nicefrac23$.
This estimate systematically yields lower velocities
because the ejecta become transparent at longer wavelengths, which reduces the Rosseland mean and places the corresponding photosphere
deeper than the region where the $B$-band continuum and the Fe$\,$\textsc{ii} lines form.
That is why the Rosseland photospheric velocity is in much worse agreement with observations.

\section{Discussion and conclusions}\label{sec:discussion}
We presented that the luminosity evolution of SN~2018ibb after $300\,\mathrm{d}$ since explosion can be well reproduced by adding the luminosity contributions from the CSM interaction. Based on the inferred CSM interaction luminosity contribution, we estimated that the PISN progenitor of SN~2018ibb had the mass-loss rate of $10^{-2}\,(v_\mathrm{CSM}/1{,}000\,\kmps)\,\Msunpyr$ in the last $28\,(v_\mathrm{CSM}/1{,}000\,\kmps)^{-1}\,\mathrm{yr}$ to explosion. The estimated mass-loss rate is, however, much higher than those expected from hot stellar winds from H-free stars \citep{Vink2022}. In addition, if such a high mass-loss rate continues for a long time, the progenitor would not have kept the large mass required for the PISN explosion. Thus, the mass-loss enhancement should have only occurred in the final, short moment before the explosion. This is consistent with the enhanced mass-loss period estimate from the light-curve modeling.

The physical mechanism of the mass-loss enhancement in the final decades towards the PISN explosion is unclear. The $121\,\Msun$ progenitor model is found to trigger C shell burning after contraction at around 30~years before explosion, matching the estimated timescale for the mass-loss enhancement when $v_\mathrm{CSM}=1{,}000~\kmps$. The core contraction with the shell burning initiates the expansion of the progenitor, which may lead to the mass-loss enhancement. Alternatively, as the progenitor evolves closer to the dynamically unstable conditions triggered by the pair creations, the progenitor gets closer to the dynamically unstable configuration that may also lead to the mass-loss enhancement. Although we assumed the wind density profile in this study, it is also possible that the CSM is formed by eruptive mass loss. Such a mass loss can also be related to binary mass transfer \citep[e.g.,][]{Ercolino2025}. Another possibility is that the mass-loss rate is actually not enhanced but the mass lost from the progenitor was accumulated near it because of, e.g., its strong magnetic fields. It has been suggested that the existence of strong magnetic fields can effectively suppress the mass-loss rate by confining the material released by the stellar wind \citep[e.g.,][]{Petit2017,Georgy2017}. Such a confined material may stay near the stellar surface, affecting the PISN light curves as presented in this paper. Further investigations are required if the magnetically confined CSM can extend close to $10^{17}\,\mathrm{cm}$.

As we start discovering PISN candidates, we start realizing that they often have signatures of unexpected CSM interaction. Another recent example is SN~2023vbw, which is suggested to be a PISN candidate showing CSM interaction signatures \citep{Hiramatsu2026}. Because the overall mass-loss rates of PISN progenitors need to be low to sustain their mass to explode as PISNe, PISN progenitors were not considered to be surrounded by dense CSM. A significant fraction of SNe from massive stars, however, are currently recognized to experience mass-loss enhancement within decades before their explosion \citep[e.g.,][]{Pastorello2007,Yaron2017,Forster2018,Ho2019,JacobsonGalan2022,Chen2026}. Therefore, it is possible that PISN progenitors also experience similar short-term enhanced mass loss just before explosion, affecting PISN observational properties. Most predicted PISN properties so far do not take the existence of dense CSM into account, but PISN models are likely required to be updated by taking the dense CSM into account. Especially, the existence of the dense CSM is likely to make PISNe bluer than previously predicted and the bluer PISNe may affect the expected discovery rates of PISNe at high redshifts \citep[e.g.,][]{Moriya2022a,Moriya2022}.

\section*{Conflict of Interest Statement}

The authors declare that the research was conducted in the absence of any commercial or financial relationships that could be construed as a potential conflict of interest.

\section*{Author Contributions}


TJM led this project and computed the numerical models.
SS initiated this project with TJM and provided the observational data of SN~2018ibb.
AH provided the \textsc{Kepler} progenitor models and contributed to their discussion in the manuscript.
SB and MP develop and support the basic version of \texttt{STELLA} code, and they produced Figure~\ref{fig:phv}.
All the authors contributed to the text.

\section*{Funding}
TJM is supported by the Grants-in-Aid for Scientific Research of the Japan Society for the Promotion of Science (JP24K00682, JP21H04997, JP24H00002, JP24H00027, JP24K00668, 26H00849) and by the Australian Research Council (ARC) through the ARC's Discovery Projects funding scheme (project DP240101786).  AH is supported by the ARC though DP240103174 and DP240101786.
SB's work was carried out within the framework of the state assignment of NRC "Kurchatov Institute".
MP is supported by the RSF grant 24-12-00141.

\section*{Acknowledgments}
We would like to thank the anonymous referees for their constructive comments that improved this manuscript.
Numerical computations were in part carried out on PC cluster at the Center for Computational Astrophysics, National Astronomical Observatory of Japan.
This work made use of \texttt{OverCite} \citep{Shariat2026}, an in-editor citation tool for \LaTeX.


\section*{Data Availability Statement}
The data underlying this article will be shared on reasonable request to the corresponding author.

\newcommand{\aj}{AJ}         
\newcommand{\araa}{ARA\&A}       
\newcommand{\apj}{ApJ}       
\newcommand{\apjl}{ApJL}       
\newcommand{\apjs}{ApJS}       
\newcommand{\apss}{Ap\&SS}       
\newcommand{\aap}{A\&A}        
\newcommand{\aapr}{A\&AR}       
\newcommand{\aaps}{A\&AS}        
\newcommand{\baas}{BAAS}       
\newcommand{\icarus}{ICARUS}      
\newcommand{\mnras}{MNRAS}       
\newcommand{\prd}{Phys.\ Rev.\ D}         
\newcommand{\prl}{Phys.\ Rev.\ Lett.}        
\newcommand{\pasp}{PASP}       
\newcommand{\pasj}{PASJ}        
\newcommand{\solphys}{Sol.\ Phys.}     
\newcommand{\ssr}{Space\ Sci.\ Rev.}         
\newcommand{\nat}{Nature}        
\newcommand{\iaucirc}{IAU\ Circ.}     
\newcommand{\gca}{Geochim.\ Cosmochim.\ Acta}        
\newcommand{\jgr}{J.\ Geophys.\ Res.}          
\newcommand{\nphysa}{Nucl.\ Phys.\ A}      
\newcommand{\procspie}{Proc.\ SPIE}   
\newcommand{\aip}{AIP Conf.\ Proc.}        
\newcommand{\asp}{ASP Conf.\ Ser.}         
\newcommand{\physrep}{PhR}

\bibliographystyle{Frontiers-Harvard} 
\bibliography{test}

\end{document}